\documentclass{article}

    \PassOptionsToPackage{numbers, compress}{natbib}

    \usepackage[preprint]{neurips_2019}

\usepackage[utf8]{inputenc} 
\usepackage[T1]{fontenc}    
\usepackage{hyperref}       
\usepackage{url}            
\usepackage{booktabs}       
\usepackage{amsfonts}       
\usepackage{nicefrac}       
\usepackage{microtype}      

\usepackage{amsmath}
\usepackage{graphicx}
\usepackage{dsfont}
\usepackage{booktabs}
\usepackage{multirow}
\usepackage{wrapfig}
\usepackage{float}
\usepackage{subfigure}
\usepackage{caption}
\usepackage{ownstyles}
\usepackage{tabularx}
\usepackage{chngcntr}

\definecolor{codegreen}{rgb}{0,0.6,0}
\definecolor{codegray}{rgb}{0.5,0.5,0.5}
\definecolor{codepurple}{rgb}{0.58,0,0.82}
\definecolor{backcolour}{rgb}{0.95,0.95,0.92}

\lstdefinestyle{mystyle}{
    backgroundcolor=\color{backcolour},   
    commentstyle=\color{codegreen},
    keywordstyle=\color{magenta},
    numberstyle=\tiny\color{codegray},
    stringstyle=\color{codepurple},
    basicstyle=\ttfamily\scriptsize,
    breakatwhitespace=false,         
    breaklines=true,                 
    captionpos=b,                    
    keepspaces=true,                 
    numbers=left,                    
    numbersep=5pt,                  
    showspaces=false,                
    showstringspaces=false,
    showtabs=false,                  
    tabsize=2
}

\newif\ifcomments

\commentstrue

\ifcomments
\newcommand{\comments}[1]{#1}
\else
\newcommand{\comments}[1]{}
\fi

\title{A Tutorial on Structural Optimization}

\author{%
  Sam Greydanus \\
  The ML Collective\\
  \texttt{greydanus.17@gmail.com} \\
}

\begin{document}

\maketitle

\begin{abstract}
  Structural optimization is a useful and interesting tool. Unfortunately, it can be hard for new researchers to get started on the topic because existing tutorials assume the reader has substantial domain knowledge. They obscure the fact that structural optimization is really quite simple, elegant, and easy to implement. With that in mind, let's write our own structural optimization code\footnote{Code is available at \texttt{https://bit.ly/394DUcL}}, from scratch, in 180 lines.
\end{abstract}

\begin{figure}[h!] \centering
\includegraphics[width=\textwidth]{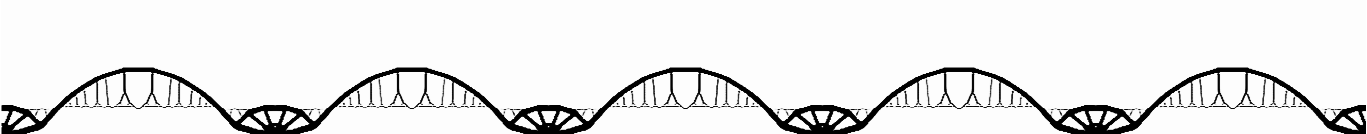}
\figlabel{fig1}
\end{figure}

\begin{lstlisting}[language=Python]
import time, nlopt                                                # for optimization
import numpy as np                                                # for dense matrix ops
import matplotlib.pyplot as plt                                   # for plotting
import autograd, autograd.core, autograd.extend, autograd.tracer  # for adjoints
import autograd.numpy as anp      
import scipy, scipy.ndimage, scipy.sparse, scipy.sparse.linalg    # sparse matrices
\end{lstlisting}

\section{Problem setup} \seclabel{intro}

The goal of structural optimization is to place material in a design space so that it rests on some fixed points or "normals" and resists a set of applied forces or \textit{loads} as efficiently as possible. To see how we might set this up, let's start with a beam design problem from \citet{andreassen2011efficient}:

\begin{figure}[h!] \centering
\includegraphics[width=.35\textwidth]{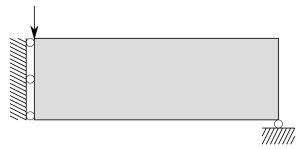}
\figlabel{fig2} \caption{The design region of an MBB beam.}
\end{figure}

The large gray rectangle here represents the design space. We are going to enforce symmetry by optimizing half of the beam and then mirroring the result around the left wall. This means that the center of the beam is actually on the left side of the diagram. This is where the load force, denoted by the downwards-pointing arrow, is being applied. There are horizontally fixed points here as well. They represent forces transmitted to this half of the beam from its other half. Meanwhile, the vertically fixed point at the bottom right corner of the design space corresponds to a normal force from some external support, perhaps the top of a wall.

\textbf{Finite elements.} Although the physics of elastic materials is continuous, our computer can only work with discrete approximations. This means that we have to cut the design space up into a discrete number of regions or \textit{finite elements} which, when allowed to interact, reproduce the behavior of an elastic solid as realistically as possible. We can link their boundaries together with a set of nodes and allow these nodes to interact with one another as though connected by springs. This way, whenever a force is applied to one node, it transmits a fraction of that force on to all the other nodes in the structure, causing each to move a small amount and, in doing so, deform the finite elements. As this happens, the entire structure deforms as though it were an elastic solid.

There are many ways to choose the arrangement of these finite elements. The simplest one is to make them square and organize them on a rectangular grid.

\begin{figure}[h!] \centering
\includegraphics[width=.35\textwidth]{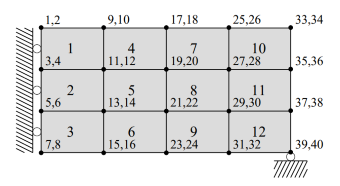}
\figlabel{fig3} \caption{Dividing the design region into finite elements.}
\end{figure}

In the diagram above, there are 12 elements with four nodes per element and two degrees of freedom (DOFs) per node. The first is horizontal and the second is vertical. The numbering scheme proceeds columnwise from left to right so that the horizontal and vertical displacements of node $n$ are given by DOFs $2n-1$ and $2n$ respectively. As the authors point out, this grid structure is useful because it can be exploited "...in order to reduce the computational effort in the optimization loop..." It also simplifies the code.

\textbf{Python representations.} Given this problem setup, every DOF in our design space can either have a force applied to it or be fixed by a normal force. For a design space that is $y$ units high and $x$ units wide, we can represent these parts of the problem setup with NumPy arrays called \texttt{forces} and \texttt{normals}, each of shape $(y+1,x+1,2)$. Here the first two axes index over all the nodes in the design space and the third axis indexes over the two DOFs available to each node. Starting with the code below – and continuing throughout the rest of this tutorial – we are going to flatten these arrays to one dimension.

There are a few other important details. The \texttt{mask} variable can be either a scalar of value 1 (no mask) or an array of shape $(x,y)$. As a default, we will use no mask. Then there are all the material constants, constraints, filter widths, and so forth to consider. For these, we use the values reported by \citet{andreassen2011efficient}. Finally, we have the \texttt{mbb\_beam} function which sets up the forces and normals particular to the MBB beam design constraints. This function can easily be swapped out if we wish to design a structure with different constraints.

\begin{lstlisting}[language=Python, firstnumber=7]
class ObjectView(object):
    def __init__(self, d): self.__dict__ = d
    
def get_args(normals, forces, density=0.4):  # Manage the problem setup parameters
  width = normals.shape[0] - 1
  height = normals.shape[1] - 1
  fixdofs = np.flatnonzero(normals.ravel())
  alldofs = np.arange(2 * (width + 1) * (height + 1))
  freedofs = np.sort(list(set(alldofs) - set(fixdofs)))
  params = {
      # material properties
      'young': 1, 'young_min': 1e-9, 'poisson': 0.3, 'g': 0,
      # constraints
      'density': density, 'xmin': 0.001, 'xmax': 1.0,
      # input parameters
      'nelx': width, 'nely': height, 'mask': 1, 'penal': 3.0, 'filter_width': 1,
      'freedofs': freedofs, 'fixdofs': fixdofs, 'forces': forces.ravel(),
      # optimization parameters
      'opt_steps': 80, 'print_every': 10}
  return ObjectView(params)

def mbb_beam(width=80, height=25, density=0.4, y=1, x=0):  # textbook beam example
  normals = np.zeros((width + 1, height + 1, 2))
  normals[-1, -1, y] = 1
  normals[0, :, x] = 1
  forces = np.zeros((width + 1, height + 1, 2))
  forces[0, 0, y] = -1
  return normals, forces, density
\end{lstlisting}

\begin{figure}[h!] \centering
\includegraphics[width=\textwidth]{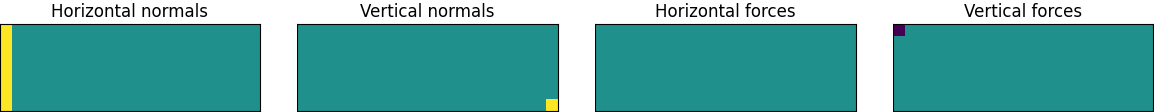}
\caption{Visualizing the normals and forces in the MBB beam setup. The rectangle represents the design area and the colored pixels represent matrix entries (blue = -1, green = 0, and yellow = 1).}
\figlabel{fig4}
\end{figure}

\vspace{.2cm}

\textbf{The density method.} Now that we have parameterized the design space, it is time to parameterize the material that moves around on it. At a high level, each finite element is going to have a certain density of material, given by some number between 0 and 1. We will use this density to determine the element stiffness coefficient $E_e$, also called Young's modulus. In the nodes-connected-by-springs analogy, this coefficient would control all the spring constants.

Let's discuss how to choose the mapping between finite element density $x_e$ and Young's modulus in more detail. First of all, we'd like to avoid having any elements with zero stiffness. When this happens, they stop transmitting forces to their neighbors before optimization is complete and we are liable to end up with suboptimal solutions. We can prevent this by giving each finite element a baseline stiffness, $E_{min}$, regardless of whether it has any material density.

We'd also like black-and-white final solutions. In other words, although our design space may start out with material densities of 0.5, by the end of optimization we'd like all of the grid cells to have densities very close to either 0 or 1. We can ensure this happens by raising our densities to a power $p$ greater than one (typically $p=3$) so as to make our structure's stiffness more sensitive to small changes in density.

Putting these ideas together, we obtain the "modified SIMP" equation from \citet{andreassen2011efficient}:

$$E_e(x_e)=E_{min} + x^p_e(E_0-E_{min}), \quad \quad x_e \in [0,1]$$

Here $E_0$ is the stiffness of the material. For a comparison between modified SIMP and other approaches, see \citet{sigmund2007topology}.

\textbf{Filtering.} Finally, in order to avoid grid-level pathologies (especially scenarios where a grid element with full density ends up next to a grid element with zero density and a discontinuity occurs), we are going to use a 2D Gaussian filter\footnote{\cite{andreassen2011efficient} use a cone filter; we found that a Gaussian filter gave similar results and was easier to implement.} to smooth the grid densities. This technique, called "filtering" shows up in almost all physics simulations where continuous fields have to be discretized.

\begin{lstlisting}[language=Python, firstnumber=35]
def young_modulus(x, e_0, e_min, p=3):
  return e_min + x ** p * (e_0 - e_min)

def physical_density(x, args, volume_contraint=False, use_filter=True):
  x = args.mask * x.reshape(args.nely, args.nelx)  # reshape from 1D to 2D
  return gaussian_filter(x, args.filter_width) if use_filter else x  # maybe filter

def mean_density(x, args, volume_contraint=False, use_filter=True):
  return anp.mean(physical_density(x, args, volume_contraint, use_filter)) / anp.mean(args.mask)
\end{lstlisting}

At this point, we have constructed a finite element parameterization of an elastic solid. We are applying forces to this solid in some places and supporting it with fixed points in others. As it deforms, it stretches and compresses in proportion to the stiffness of its finite elements. Now the question we need to ask is: \textit{what does the best structure look like under these conditions?}

\section{The objective function} \seclabel{objective-function}

At a high level, the best structure is the one that minimizes the elastic potential energy or \textit{compliance} of the 2D grid of springs. We can express this idea mathematically as follows:
\begin{align*}
\underset{\mathbf{x}}{\textrm{min}}: \quad & c(\mathbf{x})=\mathbf{U}^T\mathbf{K}\mathbf{U} = \sum_{e=1}^NE_e(x_e)\mathbf{u}_e^T\mathbf{k}_0\mathbf{u}_e \\
& \textrm{Potential energy (compliance) of a 2D grid of springs}\\
&\\
\textrm{subject to}: \quad & V(\mathbf{x})/V_0 = f \qquad \quad \textrm{A fixed quantity of material} \\
& \mathbf{0 \leq x \leq 1} \qquad \qquad ~ \textrm{Densities that remain between 0 and 1} \\
& \mathbf{KU=F}  \qquad \qquad ~~~ \textrm{Hooke's law for a 2D grid of springs} \\
\end{align*}
Here $c$ is the compliance, $\mathbf{x}$ is a vector containing the material densities of the elements, $\mathbf{K}$ is the global stiffness matrix, $\mathbf{U}$ is a vector containing the displacements of the nodes, and $E_e$ is Young's modulus. The external forces or "loads" are given by the vector $\mathbf{F}$.

We can write the core part of this objective, the part that says $c(\mathbf{x})=\mathbf{U}^T\mathbf{K}\mathbf{U}$, as a high-level objective function that calls a series of subroutines.

\begin{lstlisting}[language=Python, firstnumber=44]
def objective(x, args, volume_contraint=False, use_filter=True):
  kwargs = dict(penal=args.penal, e_min=args.young_min, e_0=args.young)
  x_phys = physical_density(x, args, volume_contraint=volume_contraint, use_filter=use_filter)
  ke     = get_stiffness_matrix(args.young, args.poisson)  # stiffness matrix
  u      = displace(x_phys, ke, args.forces, args.freedofs, args.fixdofs, **kwargs)
  c      = compliance(x_phys, u, ke, **kwargs)
  return c
\end{lstlisting}

\section{Computing sensitivities} \seclabel{sensitivities}
The objective function gives us a single number, $c(\mathbf{x})$, which we can use to rate the quality of our structure. But the question remains: \textit{how should we update $\mathbf{x}$ so as to minimize this number?} To answer this question, we need to compute the gradients or \textit{sensitivities} of $c$ with respect to $\mathbf{x}$. These sensitivities will give us the direction to move $\mathbf{x}$ in order to decrease $c$ as much as possible. Ignoring filtering for a moment and applying the chain rule to the first line of the objective function, we obtain

\begin{align}
\frac{\partial c}{\partial x_e} &= -px_e^{p-1}(E_0-E_{min})\mathbf{u}_e^T\mathbf{k}_0\mathbf{u}
\end{align}

If we want to add filtering back in, the notation becomes a bit more complicated. But we're not going to do that here because, actually, we don't need to calculate these sensitivities by hand. There is an elegant little library called Autograd\footnote{https://github.com/HIPS/autograd} which can do this for us using a process called \textit{automatic differentiation} \cite{rall1996introduction}.

\textbf{Custom gradients.} There are a few cases where we need to operate on NumPy arrays with functions from other libraries. In these cases, we need to define a custom gradient function so that Autograd knows how to differentiate through them. For example, in the code we have already written, the \texttt{gaussian\_filter} function comes from the \texttt{scipy} library. Here's how we can wrap that function to make it work properly with Autograd:

\begin{lstlisting}[language=Python, firstnumber=51]
@autograd.extend.primitive
def gaussian_filter(x, width): # 2D gaussian blur/filter
  return scipy.ndimage.gaussian_filter(x, width, mode='reflect')

def _gaussian_filter_vjp(ans, x, width): # gives the gradient of orig. function w.r.t. x
  del ans, x  # unused
  return lambda g: gaussian_filter(g, width)
autograd.extend.defvjp(gaussian_filter, _gaussian_filter_vjp)
\end{lstlisting}

\section{Implementing the physics} \seclabel{implementing-physucs}
In between $\mathbf{x}$) and $c(\mathbf{x})$, there are a series of physics functions that we still need to implement.

\textbf{Compliance.} At a high level, the compliance is just $\mathbf{U}^T\mathbf{K}\mathbf{U}$. But $\mathbf{U}$ and $\mathbf{K}$ are very sparse so it's much more efficient to calculate $\sum_{e=1}^NE_e(x_e)\mathbf{u}_e^T\mathbf{k}_0\mathbf{u}_e$. That's what we will do in the code below. It's a little hard to follow because everything is vectorized (hence the einsums) but this does speed things up considerably compared to a `for` loop.

\textbf{The element stiffness matrix.} The variable $\mathbf{k}\_0$ that appears in the compliance calculation is called the element stiffness matrix. An intuitive way to think about this matrix is as a 2D analogue of the spring constant $k$ in a simple harmonic oscillator. The reason it is a matrix (instead of a scalar or a vector) is that we need to take into account all of the various interaction terms between the corner nodes in a square finite element.\footnote{Deriving the specific entries of the element stiffness matrix takes quite a few steps. We won't go through all of them here, but you can walk yourself through them using this textbook chapter: \texttt{http://solidmechanics.org/text/Chapter7\_2/Chapter7\_2.htm.}} When we represent the displacement of all these nodes with a vector $u=[u^a_1,u^a_2,u^b_1,u^b_2,u^c_1,u^c_2,u^d_1,u^d_2]$, then it becomes easy to calculate the potential energy of the system: we just write $PE = \frac{1}{2}u^Tk_0u$ (this is the 2D analogue to the potential energy of a 1D harmonic oscillator, which is written as $\frac{1}{2}kx^2$).

From this you should be able to see why compliance is the potential energy of the entire structure: it's just a sum over the potential energies of all the finite elements. You should note that each term in the sum is getting scaled by a factor of $E_e(x_e)$. This is happening because the stiffness matrix varies with Young's modulus, and we have made Young's modulus dependent on the local material density.

\textbf{Material constants.} You'll notice that two material constants appear in the element stiffness matrix. The first is Young's modulus which measures the stiffness of a material. Intuitively, it is the distortion per unit of force ("How hard do you need to pull a rubber band to stretch it one inch?"). A more technical definition is \textit{the ratio of tensile stress to tensile strain}. The Poisson coefficient, meanwhile, measures the amount of contraction in the direction perpendicular to a region of stretching, due to that stretching ("How much thinner does the rubber band get when you stretch it one inch?"). A technical definition is \textit{the ratio between the lateral contraction per unit length and the longitudinal extension also per unit length.} Both of these coefficients come into play when we construct the element stiffness matrix.

\begin{lstlisting}[language=Python, firstnumber=59]
def compliance(x_phys, u, ke, *, penal=3, e_min=1e-9, e_0=1):
  nely, nelx = x_phys.shape
  ely, elx = anp.meshgrid(range(nely), range(nelx))  # x, y coords for the index map

  n1 = (nely+1)*(elx+0) + (ely+0)  # nodes
  n2 = (nely+1)*(elx+1) + (ely+0)
  n3 = (nely+1)*(elx+1) + (ely+1)
  n4 = (nely+1)*(elx+0) + (ely+1)
  all_ixs = anp.array([2*n1, 2*n1+1, 2*n2, 2*n2+1, 2*n3, 2*n3+1, 2*n4, 2*n4+1])
  u_selected = u[all_ixs]  # select from u matrix

  ke_u = anp.einsum('ij,jkl->ikl', ke, u_selected)  # compute x^penal * U.T @ ke @ U
  ce = anp.einsum('ijk,ijk->jk', u_selected, ke_u)
  C = young_modulus(x_phys, e_0, e_min, p=penal) * ce.T
  return anp.sum(C)

def get_stiffness_matrix(e, nu):  # e=young's modulus, nu=poisson coefficient
  k = anp.array([1/2-nu/6, 1/8+nu/8, -1/4-nu/12, -1/8+3*nu/8,
                -1/4+nu/12, -1/8-nu/8, nu/6, 1/8-3*nu/8])
  return e/(1-nu**2)*anp.array([[k[0], k[1], k[2], k[3], k[4], k[5], k[6], k[7]],
                               [k[1], k[0], k[7], k[6], k[5], k[4], k[3], k[2]],
                               [k[2], k[7], k[0], k[5], k[6], k[3], k[4], k[1]],
                               [k[3], k[6], k[5], k[0], k[7], k[2], k[1], k[4]],
                               [k[4], k[5], k[6], k[7], k[0], k[1], k[2], k[3]],
                               [k[5], k[4], k[3], k[2], k[1], k[0], k[7], k[6]],
                               [k[6], k[3], k[4], k[1], k[2], k[7], k[0], k[5]],
                               [k[7], k[2], k[1], k[4], k[3], k[6], k[5], k[0]]])
\end{lstlisting}

\textbf{Calculating displacements.} Now we need to tackle one of the most important physics problems: calculating the displacements of the nodes. The way to do this with a 1D spring would be to solve the equation $F=kx$ for $x$. Here we can do the same thing, except by solving the matrix equation $\mathbf{F=KU}$. For a system with $N$ nodes with 2 degrees of freedom each, the matrix $\mathbf{K}$ will have dimensions $2N$ x $2N$. This gives us a system of $2N$ simultaneous linear equations for $2N$ unknown node displacements.

\textbf{A global stiffness matrix with $N$ nodes.} The number of nodes  $N$ grows as the product of the width and height of our design space. Thus it is not unusual to have over $10^4$ nodes in a design space. Since the size of  $\mathbf{K}$ grows as $N^2$, it quickly becomes too large to fit in memory. For example, using $10^4$ nodes and the \texttt{np.float32} data format, we get a $\mathbf{K}$ matrix that consumes 1.6 GB of RAM. Increasing its width and height by 50\% increases that number to 8 GB. This is not a sustainable rate of growth.

Luckily, since our nodes are locally-connected, most of the entries in $\mathbf{K}$ are zero. We can save a vast amount of memory by representing it with a sparse "coordinate list" or COO format. The purpose of the \texttt{get\_k} function below is to assemble such a matrix.\footnote{If you want to see all the details for how this matrix is constructed, read the "global stiffness matrices with $N$ nodes" section of this textbook chapter: \texttt{http://solidmechanics.org/text/Chapter7\_2/Chapter7\_2.htm.}}

\textbf{The sparse matrix solve.} Having constructed $\mathbf{K}$, all we have left to do is solve the system of equations. This is the most important part of the \texttt{displace} function. It uses Scipy's \texttt{SuperLU} function (which supports COO) to solve for nodal displacements without ever instantiating a $2N$ x $2N$ matrix.

\begin{lstlisting}[language=Python, firstnumber=86]
def get_k(stiffness, ke):
  # Constructs sparse stiffness matrix k (used in the displace fn)
  # First, get position of the nodes of each element in the stiffness matrix
  nely, nelx = stiffness.shape
  ely, elx = anp.meshgrid(range(nely), range(nelx))  # x, y coords
  ely, elx = ely.reshape(-1, 1), elx.reshape(-1, 1)

  n1 = (nely+1)*(elx+0) + (ely+0)
  n2 = (nely+1)*(elx+1) + (ely+0)
  n3 = (nely+1)*(elx+1) + (ely+1)
  n4 = (nely+1)*(elx+0) + (ely+1)
  edof = anp.array([2*n1, 2*n1+1, 2*n2, 2*n2+1, 2*n3, 2*n3+1, 2*n4, 2*n4+1])
  edof = edof.T[0]
  x_list = anp.repeat(edof, 8)  # flat list pointer of each node in an element
  y_list = anp.tile(edof, 8).flatten()  # flat list pointer of each node in elem

  # make the global stiffness matrix K
  kd = stiffness.T.reshape(nelx*nely, 1, 1)
  value_list = (kd * anp.tile(ke, kd.shape)).flatten()
  return value_list, y_list, x_list

def displace(x_phys, ke, forces, freedofs, fixdofs, *, penal=3, e_min=1e-9, e_0=1):
  # Displaces the load x using finite element techniques (solve_coo=most of runtime)
  stiffness = young_modulus(x_phys, e_0, e_min, p=penal)
  k_entries, k_ylist, k_xlist = get_k(stiffness, ke)

  index_map, keep, indices = _get_dof_indices(freedofs, fixdofs, k_ylist, k_xlist)
  
  u_nonzero = solve_coo(k_entries[keep], indices, forces[freedofs], sym_pos=True)
  u_values = anp.concatenate([u_nonzero, anp.zeros(len(fixdofs))])
  return u_values[index_map]
\end{lstlisting}

\section{Sparse matrix helper functions.}

You may notice that the \texttt{displace} function uses a helper function, \texttt{\_get\_dof\_indices}, to update $\mathbf{K}$'s indices. The point here is to keep only the degrees of freedom that were actually free in the problem setup (the \texttt{freedofs}). To do this, we need to remove the degrees of freedom where normal forces were introduced (the \texttt{fixdofs}). 

The second function is the \texttt{inverse\_permutation} function. It is a mathematical operation\footnote{See \texttt{https://mathworld.wolfram.com/InversePermutation.html.}} that gives us the indices needed to undo a permutation. If \texttt{ixs} is a list of indices that permutes the list \texttt{A}, then this function gives us a second list of indices \texttt{inv\_ixs} such that \texttt{A[ixs][inv\_ixs] = A}.

\begin{lstlisting}[language=Python, firstnumber=117]
def _get_dof_indices(freedofs, fixdofs, k_xlist, k_ylist):
  index_map = inverse_permutation(anp.concatenate([freedofs, fixdofs]))
  keep = anp.isin(k_xlist, freedofs) & anp.isin(k_ylist, freedofs)
  # Now we index an indexing array that is being indexed by the indices of k
  i = index_map[k_ylist][keep]
  j = index_map[k_xlist][keep]
  return index_map, keep, anp.stack([i, j])

def inverse_permutation(indices):  # reverses an index operation
  inverse_perm = np.zeros(len(indices), dtype=anp.int64)
  inverse_perm[indices] = np.arange(len(indices), dtype=anp.int64)
  return inverse_perm
\end{lstlisting}

\textbf{Custom gradients for a sparse matrix solve.} Our sparse solve, like our 2D Gaussian filter, comes from the SciPy library and is not supported by Autograd. So we need to tell Autograd how to differentiate through it. To do this, we'll copy a few lines of code from \cite{hoyer2019neural}\footnote{GitHub permalink: \texttt{https://bit.ly/39KN66H}}.

These lines are similar to Autograd's implementation\footnote{\texttt{https://bit.ly/38502Ux}} of the gradient of a matrix solve. The main difference is that whereas the Autograd version is written for dense matrices, this version is written for sparse matrices. The underlying mathematical idea is the same either way; see \textit{An extended collection of matrix derivative results for forward and reverse mode algorithmic differentiation}\footnote{\texttt{https://bit.ly/39CXnlb}}\cite{giles2008extended} by Mike Giles for the relevant formulas.

\begin{lstlisting}[language=Python, firstnumber=129]
def _get_solver(a_entries, a_indices, size, sym_pos):
  # a is (usu.) symmetric positive; could solve 2x faster w/sksparse.cholmod.cholesky(a).solve_A
  a = scipy.sparse.coo_matrix((a_entries, a_indices), shape=(size,)*2).tocsc()
  return scipy.sparse.linalg.splu(a).solve

@autograd.primitive
def solve_coo(a_entries, a_indices, b, sym_pos=False):
  solver = _get_solver(a_entries, a_indices, b.size, sym_pos)
  return solver(b)

def grad_solve_coo_entries(ans, a_entries, a_indices, b, sym_pos=False):
  def jvp(grad_ans):
    lambda_ = solve_coo(a_entries, a_indices if sym_pos else a_indices[::-1],
                        grad_ans, sym_pos)
    i, j = a_indices
    return -lambda_[i] * ans[j]
  return jvp

autograd.extend.defvjp(solve_coo, grad_solve_coo_entries,
                       lambda: print('err: gradient undefined'),
                       lambda: print('err: gradient not implemented'))
\end{lstlisting}

And with that, we are done with the physics. Now we are ready to set up the optimization itself.

\section{Optimization}

To do this, we'll use the Method of Moving Asymptotes (MMA). Originally described by \citet{svanberg1987method} and refined in \citet{svanberg2002class}, MMA is a good fit for structural optimization problems because it accepts nonlinear inequality constraints and scales to large parameter spaces. In the code below, we rewrite the mass conservation constraint as a mass \textit{threshold} constraint so that it looks like an inequality. Then we set the density constraint by giving upper and lower bounds on the parameter space. Finally, we use Autograd to obtain gradients with respect to the objective and pass them to the solver. The NLopt package \cite{johnson2021nlopt} makes this process pretty straightforward.

Other optimization approaches we tried include optimality criteria \cite{andreassen2011efficient}, plain gradient descent, L-BFGS \cite{liu1989limited}, and the Adam optimizer \cite{kingma2014adam}. Consistent with the findings of \cite{hoyer2019neural}, MMA outperformed all of these approaches.

\begin{lstlisting}[language=Python, firstnumber=150]
def fast_stopt(args, x=None, verbose=True):
  if x is None:
    x = anp.ones((args.nely, args.nelx)) * args.density  # init mass

  reshape = lambda x: x.reshape(args.nely, args.nelx)
  objective_fn = lambda x: objective(reshape(x), args) # don't enforce mass constraint here
  constraint = lambda params: mean_density(reshape(params), args) - args.density

  def wrap_autograd_func(func, losses=None, frames=None):
    def wrapper(x, grad):
      if grad.size > 0:
        value, grad[:] = autograd.value_and_grad(func)(x)
      else:
        value = func(x)
      if losses is not None:
        losses.append(value)
      if frames is not None:
        frames.append(reshape(x).copy())
        if verbose and len(frames) % args.print_every == 0:
          print('step {}, loss {:.2e}, t={:.2f}s'.format(len(frames), value, time.time()-dt))
      return value
    return wrapper

  losses, frames = [], [] ; dt = time.time()
  print('Optimizing a problem with {} nodes'.format(len(args.forces)))
  opt = nlopt.opt(nlopt.LD_MMA, x.size)
  opt.set_lower_bounds(0.0) ; opt.set_upper_bounds(1.0)
  opt.set_min_objective(wrap_autograd_func(objective_fn, losses, frames))
  opt.add_inequality_constraint(wrap_autograd_func(constraint), 1e-8)
  opt.set_maxeval(args.opt_steps + 1)
  opt.optimize(x.flatten())
  return np.array(losses), reshape(frames[-1]), np.array(frames)
\end{lstlisting}

\section{Now we are ready to optimize our MBB beam}

\begin{lstlisting}[language=Python, firstnumber=182]
# run the simulation and visualize the result
args = get_args(*mbb_beam())
losses, x, mbb_frames = fast_stopt(args=args, verbose=True)

plt.figure(dpi=50) ; print('\nFinal design space:')
plt.imshow(x) ; plt.show()
plt.figure(dpi=100) ; print('\nFinal MBB beam design:')
plt.imshow(np.concatenate([x[:,::-1],x], axis=1)) ; plt.show()
\end{lstlisting}

\texttt{Optimizing a problem with 4212 nodes\\
step 10, loss 1.28e+03, t=1.31s\\
step 20, loss 5.38e+02, t=2.51s\\
step 30, loss 4.17e+02, t=3.92s\\
step 40, loss 3.67e+02, t=5.36s\\
step 50, loss 3.61e+02, t=6.84s\\
step 60, loss 3.58e+02, t=8.30s\\
step 70, loss 3.55e+02, t=9.67s\\
step 80, loss 3.44e+02, t=10.79s\\}
\begin{figure}[h!]
\includegraphics[width=.3\textwidth]{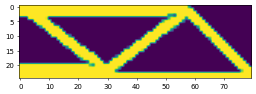}\\
\includegraphics[width=.6\textwidth]{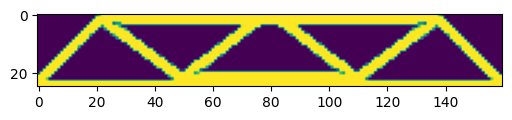}
\figlabel{fig4}
\end{figure}

\section{A few other designs}

\begin{figure}[H] \centering
\includegraphics[width=\textwidth]{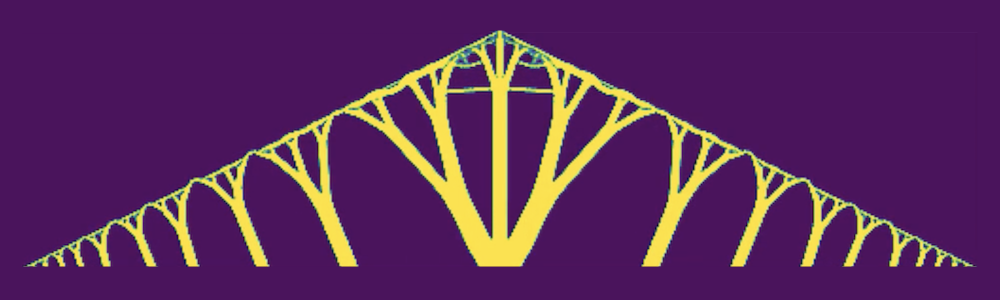}
\caption{Designing the eves of a gazebo.}
\figlabel{fig5}
\end{figure}

\begin{figure}[H] \centering
\includegraphics[width=\textwidth]{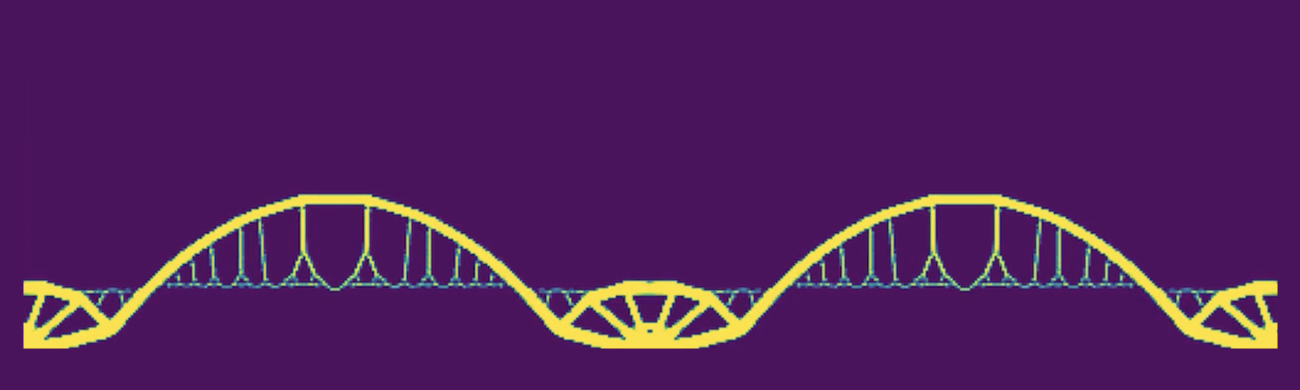}
\caption{Designing a causeway bridge.}
\figlabel{fig6}
\end{figure}

\begin{figure}[H] \centering
\includegraphics[width=.45\textwidth]{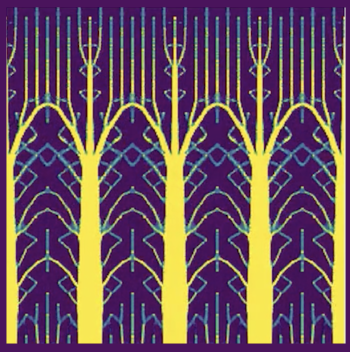}
\hspace{.25cm}
\includegraphics[width=.45\textwidth]{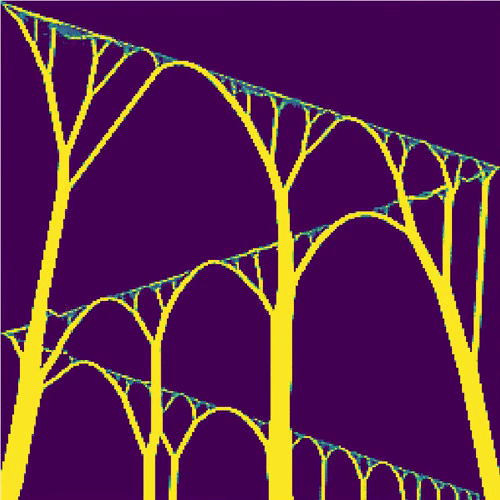}
\caption{Designing a structure that supports a grid of offset points (left) and designing a staircase (right).}
\figlabel{fig7}
\end{figure}

\begin{figure}[H] \centering
\includegraphics[width=.85\textwidth]{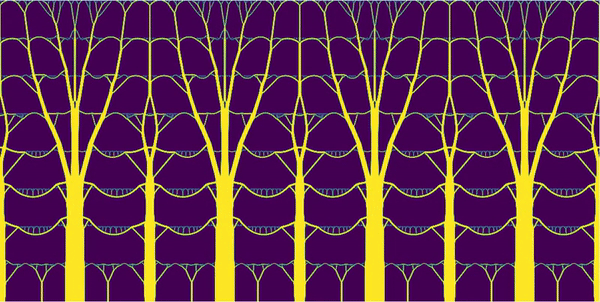}
\caption{Designing a multistory building.}
\figlabel{fig8}
\end{figure}

You can view more structures at \texttt{https://bit.ly/3KSNNYj}, courtesy of \cite{hoyer2019neural}. The problem setups are available at \texttt{https://bit.ly/3KX0uBs}.

\section{Discussion}

In sci-fi representations of healthy cities of the future, we often find manmade structures that are well integrated with their natural surroundings. Sometimes we even see a convergence where nature has adapted to the city and the city has adapted to nature. The more decadent cities, on the other hand, tend to define themselves in opposition to the patterns of nature. Their architecture is more blocky and inorganic. Perhaps tools like structural optimization can help us build the healthy cities of the future -- and steer clear of the decadent ones.

\bibliography{stopt}
\bibliographystyle{stopt}

\end{document}